\documentclass[a4paper]{spie}
\usepackage{times}
\usepackage{amsfonts}
\usepackage{amsmath}
\usepackage{amssymb}
\begin{document}
\title{The Path Integral of Feynman and ''Information Modelling'' of Processes and Systems}
\author{O.I.~Shro\supit{a}
%\email{oshro@psati.ru}
\skiplinehalf \supit{a} Volga State Academy of Telecommunications
and Informatics, Department of Information Systems and Technologies,
Chair of Software and Control of Technical Systems, 23 L. Tolstogo
str., Samara, 443011, Russian}
\authorinfo{O.I.S.: E-mail: oshro@psati.ru,  Telephone: +7 (846) 228 00 13}
\date{\today}
\maketitle

\begin{abstract}
The given article example of physical analogies to be entered
information space-time. The opportunity of Poincare group use is
shown for transition from one frame in another, for this purpose is
entered invariant velocity of transition of the information. For
calculation of information processes probability amplitudes is
offered using path integral of Feynman.
\end{abstract}

\section{\label{SecInt}INTRODUCTION}
The prominent aspect of studying of complex systems is the
opportunity of the account of all processes and the phenomena
occurring in system. The present time most studied are physical
processes. However even at studying physical processes there is a
number of problems for the decision which it is necessary to use
various models simultaneously. The examples of such problems are
descriptions of systems where thermodynamic and electromagnetic
processes simultaneously proceed. The one of ways of modelling of
such systems is the description of these systems in " information
language " i.e. when parameters of system, conditions of system are
compared to units of measure of the information, with possible
values of the information and with mechanisms of perception of the
information. Such modelling allows to abstract from consideration of
concrete processes, but it is correct to consider them in the
general behaviour of studied system.

\section{\label{SecII}The ''information modelling'' of processes and systems}
The given section of article is considered the problem so-called
"information modelling of system" for real systems, from example
physical systems. The given method of without generality restriction
can be applied for chemical, biological\cite{BaV04} and social
systems\cite{BoR05}. The as a rule, the consideration and
description of not isolated physical systems is reduced
consideration set the physical system and the physical system
environment. This is allows considering physical system as subsystem
of environment. This approach is considering problem about
interaction of two subsystems: physical systems on the one hand and
environments on the other hand. Let's consider some dynamic physical
system of not isolated from an environment. Let this subsystem is
described by some parameters set:
\begin{equation}
\begin{gathered}
\alpha\;=\;\left\{\;\alpha_{1},\;\alpha_{2},\;\ldots\;,
\;\alpha_{n-1},\;\alpha_{n}\;\right\}\>,
\end{gathered}~\label{Par-Set}
\end{equation}
here following designations are entered: the $\alpha$ -- is
designations of set parameters of consideration physical system, the
$\alpha_{i}$ -- is certain parameters of set parameters, the $n$ is
general number of parameters for consideration physical system. For
example, as such by parameter in Eq.~of~(\ref{Par-Set}) can be
chosen total moment of system, orbital moment of system, spin,
electric charge, magnetic moment, etc. The parameters set are
allowed unequivocally describing physical system. These parameters
is possible solve for functions of physical system condition, for
example, total energy of system, entropy of system, electric field
intensity, magnetic field intensity, etc. Thus are receiving
described set for the systems condition defined through set
parameters of system:
\begin{equation}
\begin{gathered}
\varphi\left(\alpha\right)\;=\;\left\{\;\varphi_{1}\left(\alpha\right),
\;\varphi_{2}\left(\alpha\right),\;\ldots\;,
\;\varphi_{m-1}\left(\alpha\right),\;\varphi_{m}\left(\alpha\right)\;\right\}\>,
\end{gathered}~\label{Fun-Set}
\end{equation}
here following designations are entered: the $\varphi$ -- is
designations of set system condition of consideration physical
systems, the $\varphi_{j}$ -- is certain system condition of set
system condition, the $m$ is general number of system condition
relation defined through set parameters of system. In case at the
system are coursing some processes or not isolated system are
influencing some actions which lead to change of parameters values
and functions of condition (which are defined through parameters of
system). These processes can be defined by some equations set,
represented through parameters and conditions of system:
\begin{equation}
\left\{
\begin{gathered}
~~~~~\psi_{1}\left[\varphi\left(\alpha\right),\;\alpha\right]\;=\;0\;,
\\[2mm]
~~~~~\psi_{2}\left[\varphi\left(\alpha\right),\;\alpha\right]\;=\;0\;,
\\
\vdots
\\
\psi_{l-1}\left[\varphi\left(\alpha\right),\;\alpha\right]\;=\;0\;,
\\[2mm]
~~~~~\psi_{l}\left[\varphi\left(\alpha\right),\;\alpha\right]\;=\;0\;\>.~\label{Eq-Set}
\end{gathered}
\right.
\end{equation}
Here $\psi_{k}$ -- is equation defined processes for system and the
$l$ is general number of processes equations. This is necessary to
note, that in among equations of processes can containing
differential equations from functions of condition and parameters.
The given equations set are representing the dynamics description,
system condition and system parameters change. The supposition is
well-known processes for physical system. However at all parameters
and possible system conditions account are necessary to face problem
of their description because of their complexity the joint account.
This especially important is considering similar of set properties
and systems description methods.

'''Information modelling'' physical system consisting that
parameters, conditions and the processes describing system, are
classified as follows:
\begin {enumerate}
\item in physical systems are allocated parameters, conditions of system,
and processes which can be interpreted, as data carriers. The given
characteristics are compared to some points of abstract information
space.
\item the values of parameters and conditions of system,
and also some processes proceeding in system, can be compared to
value of the information being in the given point of information
space.
\end {enumerate}
For an example, it is possible to consider such important
characteristic of quantum mechanical system, as spin projection of
system. The spin projection as the system characteristic can be
considered as some point in abstract information space, for example
as one bit of the information. The value of spin projection of
system can be compared to value of the information, belong in the
given point of information space. For example, positive value of a
spin projection is compared to one and negative value of spin
projection is compared to zero.

For consideration of opportunity entering 4-dimensional information
space-time and using group of Poincare transformations, необходимо
найти it is necessary to find out criterion of existence or absence
of invariant velocity of the information transitions. For example is
considered the physical system in which the electromagnetic wave
propagation with velocity of light in vacuum -- $c$ and the wave
length to equal $\lambda \; = \; 2 \; \ell_P$, where $\ell_P$ -- is
the length of Planck\footnote{The consideration of smaller length of
a wave, for example, the value of the wave length equal to value of
Planck's length $\lambda\; = \; \ell_P$, does not represent interest
if considered logic of the information bit perception. For
perception of information bit necessary to apprehend the physical
signal (wave) equal to the semi-wave length, minimally possible
value of measured which is equal to length of Planck in physical
system \cite{Shr07}.}\cite{Shr07}. That frequency of the propagation
wave is also invariant value $\nu_{p}$, this is obvious:
\begin{equation}
\nu_{p} \; = \; {{c} \over {2 \; \ell_P \>}}, \label{nuP}
\end{equation}

Let by means of the electromagnetic wave is transferred some
information, presented in the form of a binary code, for simplicity
of reasoning and without restriction of the generality. In this
frame, what transferred the information containing in one
$\lambda_{c}$-bit is necessary transferred and accepted one
semi-wave, as for recognition of one the information bit is enough
accepted or transferred the semi-wave. In the given example by
virtue of smallness the chosen the wave length conditionally is
considered the point object --- the material point is including in
the quantity and the sense of the information. On the basis of above
considered is possible made the following statement: physical
transference of the information by means of the electromagnetic wave
answers displacement of the information some bits number in
information space; otherwise this statement is made so: for moving
one bit of the information is necessary and enough moving the
semi-wave in physical space. Hence the information velocity is
proportional to frequency half of Eq.~of~(\ref{nuP}):
\begin{equation}
\nu_{c} \; = \; {{c \; \lambda_{c}} \over {\ell_P}} \>,
\label{nuPtoC}
\end{equation}
If used formula for length Planck through fundamental physical
constants: velocity of light in vacuum -- $c$, Planck's constant --
$\hbar$ the constant of gravitational interaction -- $G$:
\begin{equation}
\displaystyle{\ell_P} \; = \; \displaystyle{\sqrt {{{\hbar \;G_{N}}
\over {c^{3}}}}} \>, \label{Lp}
\end{equation}
The having substitution the Eq.~of~(\ref{Lp}) in
Eq.~of~(\ref{nuPtoC}), gives the following equation for invariant
value of information velocity:
\begin{equation}
\begin{gathered}
\displaystyle{\nu_{c}} \; = \; \displaystyle{\sqrt{{{\lambda^{2}_{c}
\; c^{5}} \over {\hbar \; G_{N}}}}} \>,
\end{gathered}
\label{Cvel}
\end{equation}

In the considered example all parameters of a transferred
electromagnetic wave were invariant, in any physical frame of
reference. By virtue of it is approved: the length of a semi-wave is
related with the information transferring in information space,
transfer with invariant velocity Eq.~of~(\ref{Cvel}).

\section{\label{SecIII}Property of Information space}
Let's note only, that position of the information should be
described by means of covariant (or contravariant) the 4-vector of
position -- $x^{\alpha}$ and the corresponding 4-vector of velocity
$v^{\alpha}$. The 4-vector should be transformed with usage of
Poincare group realization in which as invariant velocity is used
invariant value of information velocity -- $\nu_{c}$
Eq.~of~(\ref{Cvel}), at transition from one frame of reference to
other frame of reference\cite{Shr07}:
\begin{equation}
x^{\;\alpha} \; = \; \left( x^{0},\;
x^{1},\;x^{2},\;x^{3}\right)\;\>.~\label{chi}
\end{equation}
The relativistic invariance consists in scalar product preservation
at Poincare group transformations. At transition from one frame of
reference in other frame of reference of the position 4-vector
should be transformed on Poincare group representation, according
has following obvious form for a 4-vector displacement
transformation:
\begin{equation}
x^{\prime\;\alpha} \; =\; \Lambda^{\alpha}_{~\beta}\; x^{\;\beta}\;
+\;b^{\;\alpha}\>, ~\label{Pua}
\end{equation}
where $\Lambda^{\alpha}_{\beta}$ --- is Lorentz's transformation
matrix, $b^{\;\alpha}$ --- is  4-vector of time-space displacement
in information space.

In frame of if product of the 4-vector most on itself will turn out
the following equality for its component, following of requirements
of relativistic invariance is considered:
\begin{equation}
x^{2}\;=\;x^{\;\beta}\;x_{\;\beta}\; = \; \left(x^0\right)^{2}\;-\;
\left(x^1\right)^{2}\;-\; \left(x^2\right)^{2}\;-\;
\left(x^3\right)^{2} = \; \nu^{2}_{c} \; \tau^{2}\>,~\label{invX}
\end{equation}
where $x^{0}\;=\; \nu_{c}\; t$ --- is lights bit; $t$ --- is time in
the laboratory system of reference, in conformity coincides in due
course in physical system; $\nu_{c} \; \tau$
--- is the 4-interval, invariant value; $\tau$
--- is intrinsic time which, according to
identically coincides with intrinsic time in considered physical
system. From the received obvious kind of a square of the 4-interval
in this frame have relation between infinitesimal time intervals in
intrinsic system of references and laboratory system of references
\cite{Shr07}:
\begin{equation}
\nu^{2}_{c}\;\left(d\;\tau\right)^2\;= \;\nu^{2}_{c}\;
\left(d\;t\right)^2\; + \; \left(d\;x^{1}\right)^2\;+ \;
\left(d\;x^{2}\right)^2\;+ \;
\left(d\;x^{3}\right)^2\>.~\label{DintX}
\end{equation}

Let's consider the problem on displacement in information space-time
and calculation of its mean value. For this purpose we shall
consider mean displacement between two points:
\begin{equation}
\langle\Delta x^{\alpha}\rangle \; = \;
\langle\displaystyle{\left(\;{x_{b}^{\;\alpha} -
x_{a}^{\;\alpha}}\;\right)}\rangle\>,~\label{V3}
\end{equation}
Without restriction of a generality have considered displacement of
the information along one axis, for example is axis of $x^1 $. The
considering 4-velocity vector component at obvious form define as
the derivative of the information position 4-vector on intrinsic
time \cite{Shr07}:
\begin{equation}
\begin{gathered}
v^{\alpha} \; = \;{{\displaystyle{d\;x^{\alpha}}} \over
{\displaystyle{d\;\tau}}}\; = \; {\displaystyle{\displaystyle{1}}
\over
\displaystyle{{\sqrt{1\;-\;\left(\displaystyle{v/{\nu_{c}}}\right)^{2}}}}}
\; \left(v^{0}, \; v^{1},\;v^{2},\;v^{3}\right)\>,~~
v\;=\;\sqrt{\left(v^{1}\right)^{2}\; + \;\left(v^{2}\right)^{2}\; +
\;\left(v^{3}\right)^{2}}\>,~\label{vel}
\end{gathered}
\end{equation}
and enter the following dimensionless 4-vector which components are
the  4-velocity attitude invariant value of information velocity
Eq.~of~(\ref{Cvel}):
\begin{equation}
\begin{gathered}
\beta^{\;\alpha}\; = \;{\displaystyle{\displaystyle{v^{\,\alpha}}}
\over \displaystyle{\nu_{c}}}\; = \;{\displaystyle{\displaystyle{1}}
\over \displaystyle{{\sqrt{1\;-\;\beta^2}}}} \; \left( 1, \;
\beta^{1},\;\beta^{\,2},\;\beta^{\,3}\right)\>,~~~~
\displaystyle{\beta^{\;i}}=\displaystyle{{v^{\;i}}/{\nu_{c}}}\>.
\end{gathered}~\label{Bvel}
\end{equation}
The velocity of displacement of the information along this axis, at
the relativistic description have designated for convenience $u^1$:
\begin{equation}
u^1\;=\;\nu_{c}\;{{\beta^{\;1}}\over\displaystyle{{\sqrt{1\;-\;\beta^{\;2}}}}}\>.
~\label{TVel1}
\end{equation}
The displacement from a point $x_{a}^1$ in the point $x_{b}^1$
should be expressed by following integral from velocity along the
axis $x^1$:
\begin{equation}
\Delta\;
x^1\;=\;x^1_{b}-x^1_{a}\;=\;\int\limits_{x_{a}^{1}}^{\;x_{b}^{1}}u^1\;
d\;\tau\;=\;\nu_{c}\;\int\limits_{x_{a}^{1}}^{\;x_{b}^{1}}
{{\beta^{\;1}}\over\displaystyle{{\sqrt{1\;-\;\beta^{\;2}}}}}\;d\;\tau
\>.~\label{TBias1}
\end{equation}
The considering 4-velocities scalar product obvious form:
\begin{equation}
v^{\alpha}\;v_{\alpha}\; = \; {\displaystyle{\displaystyle{1}} \over
\displaystyle{{{1-\left(\displaystyle{v/{\nu_{c}}}\right)^{2}}}}}\;\left[
\left(v^0\right)^{2}-\; \left(v^1\right)^{2}\;-\;
\left(v^2\right)^{2}\;-\; \left(v^3\right)^{2}\right]\;
\equiv\;{\nu_{c}}^2\>,~\label{Pvel}
\end{equation}
and 4-velocity vector component  Eq.~of~(\ref{vel}) at statement of
Eq.~of~(\ref{TBias1}) is received following integral:
\begin{equation}
\Delta\; x^1\;=\;\int\limits_{x_{a}^{1}}^{\;x_{b}^{1}}v^1\;
d\;t\>,~\label{TBias2}
\end{equation}
where $t$-- is time in a laboratory frame of reference. The
continuous change of component of the displacement vector  along the
axis $x^1$ should be casual parameter, therefore measured
displacement should be mean displacement calculated on rules of
calculation of a continuous random variable, not concretizing the
form of the random variable probability density (to the question of
calculation of density should be return later, considering
considering path integrals of Feynman), should be calculate an mean
displacement:
\begin{equation}
\langle\Delta\;
x^1\rangle\;=\;\langle\;\int\limits_{x_{a}^{1}}^{\;x_{b}^{1}}v^1\;
d\;t\;\rangle\>.~\label{TAver1}
\end{equation}
By virtue the measured invariant mean displacement value is the
value $\lambda_{c}\;=\;$ 1 bit, therefore any mean displacement
along any space axis in considered information space should be
proportional to the integer of bits. If to assume is received
contradiction, that $\lambda_{c}$ is not invariant value that
contradicts. Therefore is received, that any mean displacement is
proportional to the integers of bits:
\begin{equation}
\langle\Delta x^1\rangle\;=\;N_{1}\;\lambda_{c}\>.~\label{TAver2}
\end{equation}
The analogously is possible to prove validity of the statement about
proportionality of components to the integer of bits for the
remained space components of the mean displacement vector. Thus
consider, that equalities for all space coordinates are already
proved by us. So, should be calculate mean value from time component
of a 4-vector of displacement:
\begin{equation}
\begin{gathered}
\Delta\; x^0\;=\;x_{b}^{0}-x_{a}^{0}\;\;=\;
\nu_{c}\;\int\limits_{x_{a}^{0}}^{\;x_{b}^{0}}{{1}\over\displaystyle{{\sqrt{1\;-\;\beta^{\,
2}}}}}\;d\;\tau\>. ~~~ \Delta\;
x^0\;=\;\nu_{c}\;\left[t_{b}-t_{a}\right]\;=\;\nu_{c}\;\Delta t \>.
\end{gathered}~\label{TTime1}
\end{equation}
The mean time interval should be received as random value:
\begin{equation}
\langle\Delta x^0\rangle\;=\;\nu_{c}\;\langle\Delta t\rangle
\>.~\label{TTave1}
\end{equation}
For calculation of mean values in the left part uses relate between
time intervals in laboratory and intrinsic frame of references (this
frames of references is considering inertial frame of references)
\cite{Shr07}:
\begin{equation}
\Delta \tau\;=\;\Delta t\;\sqrt{1\;-\;\beta^{\,2}}\>,~\label{Ttau}
\end{equation}
where $\beta$ is expressed in velocity of movement of inertial frame
of references readout concerning laboratory, as the result of
averaging the given equation and substitution of result:
\begin{equation}
\langle\Delta t\rangle\;=\;{{\lambda_{c}}\over{\sqrt{\langle
v^2\rangle}}}\;\sqrt{N_{\, 1}^2\;+ \;N_{\, 2}^2\;+ \;N_{\,
3}^2}\>.~\label{Tinter3}
\end{equation}
The Eq.~of~(\ref{Tinter3}) coincides with equation for time
component. Thus, following parities turn out for a component of
4-dimensional displacement in information space:
\begin{equation}
\begin{gathered}
N_{1} \; = \; \displaystyle{\nu_{c}\over\lambda_{c}}\;
\langle\displaystyle{\Delta x^{1}}\rangle\>,~~~N_{2} \; = \;
\displaystyle{\nu_{c}\over\lambda_{c}}\; \langle\displaystyle{\Delta
x^{2}}\rangle\>,~~~N_{3} \; = \;
\displaystyle{\nu_{c}\over\lambda_{c}}\; \langle\displaystyle{\Delta
x^{3}}\rangle\>, ~~~ \langle\Delta t\rangle\; =
\displaystyle{\lambda_{c}^{2}\over \sqrt{\langle v^2\rangle}}\;
\sqrt{N^{\,2}_{1}\; + \; N^{\,2}_{2} \; + \;
N^{\,2}_{3}}\;\>.~\label{Vxyz}
\end{gathered}
\end{equation}

\section{\label{SecI}The path integral of Feynman}
The probably to consider the question on information transference.
The information transference has chance quantity, i.e. the
information position in the given point $x^{\alpha}_{b}$ can be
found with some probability of initial point $x^{\alpha}_{a}$
transition:
\begin{equation}
\Omega\left(x^{\;\alpha}_{b},\;x^{\;\alpha}_{a}\right)=\left|
K\left(x^{\;\alpha}_{b},\;x^{\;\alpha}_{a}\right)\right|^2=
K\left(x^{\;\alpha}_{b},\;x^{\;\alpha}_{a}\right)
K^{\ast}\left(x^{\;\alpha}_{b},\;x^{\;\alpha}_{a}\right)\>,~\label{Ver}
\end{equation}
where $K\left(b,\; a\right)$ -- is generally complex amplitude of
transition from a point $x^{\;\alpha}_{a}$ to a point
$x^{\;\alpha}_{b}$, $K^{\ast}\left(b,\; a\right)$ -- is complexly
interfaced amplitude. For transition amplitude calculation has used
method path integral \cite{FeV00}. Thus, action for transference
information is defined through a 4-vector of momentum:
\begin{equation}
S_{0}\left(x^{\;\alpha}_{b},\;x^{\;\alpha}_{a}\right)=
\int\limits_{\displaystyle{x^{\;\alpha}_{a}}}^{\displaystyle{x^{\;\alpha}_{b}}}\mathfrak{L}_{\;0}\;d\tau
=
-\int\limits_{\displaystyle{x^{\alpha}_{a}}}^{\displaystyle{x^{\;\alpha}_{b}}}
p^{\;\,\alpha}\;d x_{\,\alpha}\>.~\label{S0}
\end{equation}
The transition amplitude between points $x^{\;\alpha}_{a}$ and
$x^{\;\alpha}_{b}$ was possible to calculate as integral on all
possible paths from action function, where as a constant has been
used the certain value of information constant $\hbar_{c}$
\cite{Shr07}:
\begin{equation}
K\left(x^{\;\alpha}_{b},\;x^{\;\alpha}_{a}\right)
=\displaystyle{{1}\over {N_{\infty}}}\int\limits_{\Gamma}
\exp\left[{\displaystyle{-{{i}}\;
S_{0}\left(x^{\;\alpha}_{b},\;x^{\;\alpha}_{a}\right)}}
\right]\prod\limits_{\displaystyle{x_{\;i}^{\;\alpha}}}
\displaystyle{{d^{\;4} p\left(x_{\;i}^{\;\alpha}\right)\;d^{\;4}
x\left(x_{\;i}^{\;\alpha}\right)}\over{8\; \pi^4}}\>,~\label{Kdef}
\end{equation}
where $\Gamma$ -- is a of phase space volume in which there is the
information transference on all possible paths; $x_{\;i}^{\;\alpha}$
-- is carried out $i$-partition number of information position
4-vector in information space, partition trajectory which moving the
information from initial point $x^{\;\alpha}_{a}$ in finite point
$x^{\alpha}_{b}$, it essentially depends on partition paths;
$d^{\;4} p\left(x_{i}^{\;\alpha}\right)$ and $d^{\;4}
x\left(x_{i}^{\;\alpha}\right)$ -- is corresponding $i$-partition
number to that splitting of phase 4-volume paths elements in
momentum and coordinate spaces; $N_{\infty}$ -- is normalizing
constant which can be defined from normalizing condition on unit of
probability in all information space:
\begin{equation}
\begin{gathered}
\left|\displaystyle{{1}\over
{N_{\infty}}}\;K\left(\Gamma_{\infty}\right)\right|^{2}\;=\;
\displaystyle{{1}\over {N_{\infty}
N^{\ast}_{\infty}}}\;K\left(\Gamma_{\infty}\right)K^{\ast}\left(\Gamma_{\infty}\right)\;
=\;1\>,
\end{gathered}~\label{Normdef}
\end{equation}
where $\Gamma_{\infty}$ -- is all 4-volume in information space of
the given information extends on all possible paths. The obvious
form of paths integral have calculated between two points
$x^{\;\alpha}_{a}$ and $x^{\;\alpha}_{b}$ Eq.~of~(\ref{Kdef}) which
defines amplitude of information transition, and then obvious form
of normalizing constant have calculate for free information
transference frame of Eq.~of~(\ref{Normdef}). The calculation of
paths integral is spent by partition each path into such sites,
where $p^{\;\alpha}\left(x_{\;i}^{\;\alpha}\right)$ -- is constant
everyone $i$-partition number,
$x^{\;\alpha}\left(x_{\;i}^{\;\alpha}\right)$ -- is varies linearly,
the $M$ parts number for each of paths will depend on away
partition, nevertheless, in a limit at $M\;\rightarrow\;\infty$ the
following obvious forms is received for path integral:
\begin{equation}
K\left(b,\; a\right) =\displaystyle{{1}\over
{N_{\infty}}}\int\limits_{\Gamma} \exp\left[{\displaystyle{-{{i}}\;
p_{\;\alpha}\left(x^{\alpha}_{b}\;-\;x^{\alpha}_{a}\right)}}\right]\;
\displaystyle{{d^{\;4} p}\over{8\; \pi^4}}\>.~\label{Kint}
\end{equation}
The integral of Eq.~of~(\ref{Kint}) will be possible calculate,
knowing an obvious form of 4-vector of momentum $p^{\;\alpha}$. For
calculation of mean value is necessary know in obvious form
4-density of the information transition probability from initial
point in final point. For this purpose have considered probability
of such transition Eq.~of~(\ref{Ver}) and have find density of
probability in coordinate space:
\begin{equation}
\begin{aligned}
\rho\left(x^{\;\alpha}_{b},\;x^{\;\alpha}_{a}\right)\;=\;
\displaystyle{{d\;\Omega\left(x^{\;\alpha}_{b},\;x^{\;\alpha}_{a}\right)}\over{d^{\;4}
x}}\;=\;
\displaystyle{{d\;K\left(x^{\;\alpha}_{b},\;x^{\;\alpha}_{a}\right)}\over{d^{\;4}
x}}K^{\ast}\left(x^{\;\alpha}_{b},\;x^{\;\alpha}_{a}\right)+
K\left(x^{\;\alpha}_{b},\;x^{\;\alpha}_{a}\right)
\displaystyle{{d\;K^{\ast}\left(x^{\;\alpha}_{b},\;x^{\;\alpha}_{a}\right)}\over{d^{\;4}
x}}\>.
\end{aligned}~\label{Rodef}
\end{equation}
The received definition of probability density can be used at
calculation of mean from observable values: 4-vectors of information
displacement, a 4-vector of momentum, etc. Also the given equation
Eq.~of~(\ref{Rodef}) can be used for updating model parameters in
requirements, i.e. at calculation of mean displacement in
information space:
\begin{equation}
\langle \left(x^{\;\alpha}_{b}\;-\;x^{\;\alpha}_{a}\right)\rangle =
\int\limits_{\Gamma}
\left(x^{\;\alpha}_{b}\;-\;x^{\;\alpha}_{a}\right)
\rho\left(x^{\;\alpha}_{b},\;x^{\;\alpha}_{a}\right)\;{d^{\;4} x}\>.
~\label{Delxdef}
\end{equation}

\section{\label{SecOut}Conclusions}
The considered direction of modelling of information space can be
applied without the generality limitations to the description of any
real systems for which studying information making system is
important and thus the account of processes of the various nature,
for example, physical and sociological processes is required. The
given direction of information space modelling  allows to consider,
on the one hand, objective parameters of the information, on the
other hand, allows to consider within the limits of the approach
subjective features of perception of the information. The important
question at modelling information space is a correct comparison of
the phenomena and processes proceeding in physical (chemical,
biological, social, etc.) system with three-dimensional coordinates
of information space. As a whole, apparently to the author, the
given relativistic direction of the modelling description of
information space and information systems has greater prospects from
the point of view of the description of information processes in one
language and without dependence from type of considered system or
especially their combinations.

% ****************************************************

\acknowledgments {The author grateful to A.V.Gorokhov., A.F.Krutov
and A.A. Mironov for attention to article and for numerous and
fruitful discussions.}

\end{document}